\documentclass[aps,prl,twocolumn,showpacs,preprintnumbers,amsmath,amssymb,superscriptaddress]{revtex4}%

\usepackage{graphicx}%
\usepackage{dcolumn}
\usepackage{amsmath}
\usepackage{color}
\usepackage{bm}

\begin{document}

\preprint{PREPRINT (\today)}

\title{Universal observation of multiple order parameters in cuprate superconductors }

\author{R.~Khasanov}
 \affiliation{Physik-Institut der Universit\"{a}t Z\"{u}rich,
Winterthurerstrasse 190, CH-8057 Z\"urich, Switzerland}
\author{S.~Str\"assle}
 \affiliation{Physik-Institut der Universit\"{a}t Z\"{u}rich,
Winterthurerstrasse 190, CH-8057 Z\"urich, Switzerland} %
\author{D.~Di~Castro}
 \affiliation{Physik-Institut der Universit\"{a}t Z\"{u}rich,
Winterthurerstrasse 190, CH-8057 Z\"urich, Switzerland}
\affiliation{CRS Coherentia, CNR-INFM and Dipartimento di Fisica,
Universita' di Roma "La Sapienza", P.le A. Moro 2, I-00185 Roma,
Italy}
\author{T.~Masui}
 \affiliation{Department of Physics, Osaka
University, Machikaneyama 1-1, Toyonaka, Osaka 560-0043, Japan}
\author{S.~Miyasaka}
 \affiliation{Department of Physics, Osaka
University, Machikaneyama 1-1, Toyonaka, Osaka 560-0043, Japan}
\author{S.~Tajima}
 \affiliation{Department of Physics, Osaka
University, Machikaneyama 1-1, Toyonaka, Osaka 560-0043, Japan}
\author{A.~Bussmann-Holder}
\affiliation{Max-Planck-Institut f\"ur Festk\"orperforschung,
Heisenbergstrasse 1, D-70569 Stuttgart, Germany}
\author{H.~Keller}
\affiliation{Physik-Institut der Universit\"{a}t Z\"{u}rich,
Winterthurerstrasse 190, CH-8057 Z\"urich, Switzerland}
%

\begin{abstract}
The temperature dependence of the London penetration depth
$\lambda $ was measured for an untwined single crystal of
YBa$_2$Cu$_3$O$_{7-\delta}$ along the three principal
crystallographic  directions ($a$, $b$, and $c$). Both in-plane
components ($\lambda_a$ and $\lambda_b$) show an inflection point
in their temperature dependence which is absent in the component
along the $c$~direction ($\lambda_c$). The data provide convincing
evidence that the in-plane superconducting order parameter is a
mixture of $s+d-$wave symmetry whereas it is exclusively $s-$wave
along the $c$~direction. In conjunction with previous results it
is concluded that coupled $s+d-$order parameters are universal and
intrinsic to cuprate superconductors.
\end{abstract}

\maketitle

It is believed that the CuO$_2$ planes are the essential building
units in cuprate high-temperature superconductors (HTS's)
where superconductivity occurs.
Even though either static or dynamic distortions of these planes
destroy the cubic symmetry, many theoretical approaches ignore the
observed orthorhombicity and idealize the planar structure, mostly
in order to justify a pure $d-$wave order parameter. Early on it
was, however, emphasized that cuprates must have a more complex
order parameter than just $d$-wave \cite{Muller95,Muller97},
supported by many experiments like nuclear magnetic resonance (NMR)
\cite{Martindale98}, Raman scattering \cite{Masui03,Friedl90},
angle-resolved electron tunneling \cite{Smilde05}, Andreev
reflection \cite{Deutscher05},
angular-resolved photoemission (ARPES) \cite{Lu01},
muon-spin rotation ($\mu$SR)
\cite{Khasanov07_La214,Khasanov07_Y124,Khasano07_La214_book}, and
neutron crystal-field spectroscopy \cite{Furrer07}. In addition,
experiments along the $c$~axis like, {\it e.g.}, tunneling
\cite{Sun94}, bi-crystal twist Josephson junctions \cite{Li99},
optical pulsed probe \cite{Kabanov99}, and optical reflectivity
\cite{Muller04} suggest that a pure $s-$wave order parameter is
realized here.

Multiple order parameter scenarios were  proposed shortly after the
BCS theory in order to account for a complex band structure and
interband scattering \cite{Suhl59,Moskalenko59,Kresin73}. This
approach has the advantage that high-temperature superconductivity
can easily be realized even within weak coupling theories since
interband scattering provides a pairwise exchange between different
bands which strongly enhances the transition temperature ($T_c$) as
compared to a single band model.
%
%
The first realization of two-band superconductivity has been made in
Nb doped SrTiO$_3$ \cite{Binnig80} and has not attracted very much
attention. With the discovery of high-temperature superconductivity
in MgB$_2$ two-gap superconductivity became more prominent, and
meanwhile many more systems exhibiting multi-band superconductivity
have been discovered (see, e.g.,
Refs.~\cite{Binnig80,Giubileo01,Boaknin03,Shulga98,Seyfarth05}).
Interestingly, in all these systems the coupled superconducting
order parameters are of the same symmetry, i.e., $s+s$, $d+d$. In
this respect HTS's are novel since here mixed order parameter
symmetries are realized, namely, $s+d$. Theoretically it has been
shown that mixed order parameters support even high values of $T_c$
and lead to an almost doubling of the transition temperature as
compared to coupled order parameters of the same symmetry
\cite{Bussmann-Holder03a}.

In order to prove that complex order parameters are intrinsic and
universal to HTS, previous $\mu$SR measurements
\cite{Khasanov07_La214,Khasanov07_Y124,Khasano07_La214_book} were
continued for another HTS family, namely
YBa$_2$Cu$_3$O$_{7-\delta}$.
The $\mu$SR technique has the advantage that it is bulk sensitive
and a direct probe of the London penetration depth which is highly
anisotropic in HTS's. Recent results for La$_{1-x}$Sr$_x$CuO$_4$ and
YBa$_2$Cu$_4$O$_8$
\cite{Khasanov07_La214,Khasanov07_Y124,Khasano07_La214_book} clearly
demonstrate the existence of two coupled $s+d-$wave gaps in the
CuO$_2$ planes and an $s-$wave gap along the $c$~axis in
YBa$_2$Cu$_4$O$_8$. While these findings already suggest that a
complex gap structure is intrinsic to HTS, the new results on
YBa$_2$Cu$_3$O$_{7-\delta}$, presented below, support this
conclusion consistently. We are thus reasoning that $s+d-$wave
superconductivity in the planes and $s-$wave superconductivity along
the $c$~direction are intrinsic and universal to this complex
material class. In addition, this finding imposes serious
constraints for theoretical models, {\it e.g.}, neither purely
electronic nor 2D based approaches capture this complicated picture.

The crystal was grown by a crystal pulling technique
\cite{Y123_sample_preparation} and exhibited a rectangular shape of
approximate size of 4x4x1~mm$^3$. The sample was detwinned by
annealing it under stress for 2 months at 400$^{\rm o}$C. The total
fraction, where $a$ and $b$ axis are exchanged, occupies
approximately 8 to 10\% of the entire crystal, as confirmed by
measurements with a polarized microscope. $T_c$ and the transition
width were determined by DC-magnetization measurements and found to
be 91.2~K and 2~K, respectively.

The transverse-field $\mu$SR experiments were carried out at the
$\pi$M3 beam line at the Paul Scherrer Institute (Villigen,
Switzerland). The samples were field cooled from above $T_c$ to
1.7~K in magnetic fields ranging from 0.012~T to 0.64~T. The typical
counting statistics were $\sim$20-25 million muon detections per
experimental point. The experimental data were analyzed within the
same scheme as described in
Refs.~\onlinecite{Khasanov07_La214,Khasanov07_Y124,Khasano07_La214_book}.
This is based on a four component Gaussian fit of the $\mu$SR time
spectra where one component describes the background signal stemming
from muons stopped outside the sample, and the three other
components describe the asymmetric local magnetic field $P(B)$
distribution in the superconductor in the mixed state. The magnetic
field penetration depth $\lambda$ was derived from the second moment
of $P(B)$ since $\lambda^{-4}\propto\langle\Delta
B^2\rangle\propto\sigma_{sc}^2$ \cite{Brandt88}. The superconducting
part of the square root of the second moment
($\sigma_{sc}\propto\lambda^{-2}$ ) was obtained by subtracting the
normal state nuclear moment contribution ($\sigma_{nm}$) from the
measured $\sigma$, as $\sigma_{sc}^2=\sigma^2-\sigma_{nm }^2$ (see
Ref.~\onlinecite{Khasanov07_La214} for details).

\begin{figure}[htb]
\includegraphics[width=0.8\linewidth]{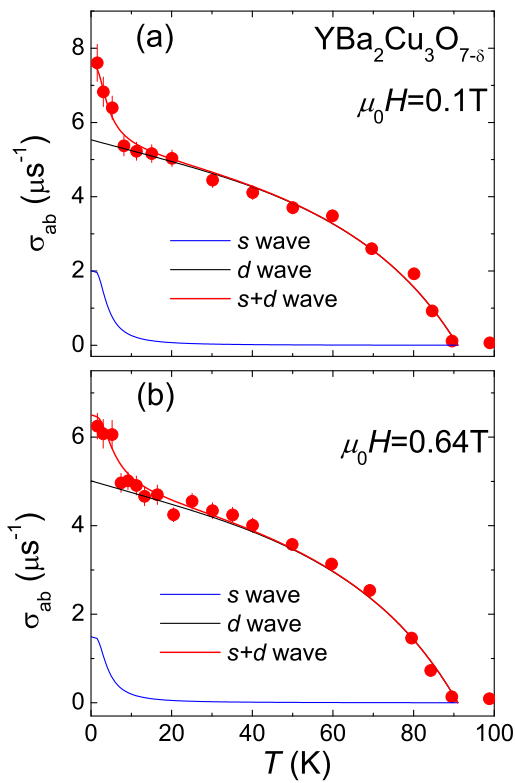}
%
\caption{(Color online) Temperature dependences of
$\sigma_{ab}\propto \lambda^{-2}_{ab}$  of
YBa$_2$Cu$_3$O$_{7-\delta}$ measured after field cooling in
$\mu_0H=$0.1~T (a) and 0.64~T (b). The red lines represent results
of a numerical calculation using the two-gap model
\cite{Khasanov07_La214,Khasanov07_Y124,Khasano07_La214_book} with
parameters as summarized in Table~\ref{Table:two-gap}. The
contributions of the small $s-$wave  and the large $d-$wave gap to
the in-plane superfluid density are shown by the blue and the black
lines, respectively.}
 \label{fig:sigma_vs_field}
\end{figure}

Since cuprates are highly anisotropic, the relation
$\sigma_{sc}\propto\lambda^{-2}$ has to be extended to account for
magnetic fields applied along the three crystallographic directions
($i,j,k=a,b,c$). For the field applied along the $i-$th principal
axis the penetration depth is determined from the second moment like
$\lambda_{jk}^{-2}=(\lambda_j\lambda_k)^{-1}\propto\sigma_{jk} =
\sqrt{\sigma_j\sigma_k}$ \cite{Thiemann89}. Here the index {\it sc}
was omitted for clarity. The magnetic field dependence of the
in-plane penetration depth has first been measured for different
fields (0.05~T, 0.1~T, 0.2~T, and 0.64~T). The field was applied
along the crystallographic $c$~axis and, subsequently, the sample
was cooled down from above $T_c$ to 1.7~K. For this field
configuration the in-plane component of the superfluid density
$\sigma_{ab}\propto\lambda^{-2}_{ab}=(\lambda_a\lambda_b)^{-1}$ was
obtained. In Fig.~\ref{fig:sigma_vs_field} $\sigma_{ab}$ is shown as
a function of temperature for two representative fields of 0.1~T and
0.64~T. In analogy to previous results
\cite{Khasanov07_La214,Khasanov07_Y124,Khasano07_La214_book}, an
inflection point in $\sigma_{ab}(T)$  is observed at $T\simeq 10$~K,
which is a typical signature for the coexistence of a small $s-$wave
and a large $d-$wave gap. Accordingly, the analysis of the data was
performed by assuming that $\sigma(T)$ can be decomposed into two
components having $d-$wave and $s-$wave symmetry as
$\sigma(T)=\sigma^s(T)+\sigma^d(T)$
\cite{Khasanov07_La214,Khasanov07_Y124,Khasano07_La214_book}. The
temperature dependences of the individual components were obtained
within the same framework as presented in
Refs.~\onlinecite{Khasanov07_La214,Khasanov07_Y124,Khasano07_La214_book}.
The comparison of experimental and theoretical results is made in
Fig.~\ref{fig:sigma_vs_field}, where the red lines refer to the sum
of the two components, whereas the blue and the black lines display
the individual $s-$ and $d-$wave contributions, respectively. For
all magnetic fields the analysis was based on common
zero-temperature gap values ($\Delta^s_0$ and $\Delta^d_0$) but
field dependent second moments [$\sigma^s(0)$, $\sigma^d(0)$]. The
zero-temperature gap values obtained in this way are
$\Delta^s_0=0.707(11)$~meV, $\Delta_0^d=22.92(9)$~meV, and are in
good agreement with results from tunneling experiments for the
$d-$wave gap (see, {\it e.g.}, Ref.~\onlinecite{Maggio-Aprile95}).
The parameters obtained from the analysis are summarized in
Table~\ref{Table:two-gap}.

\begin{figure}[htb]
\includegraphics[width=0.8\linewidth]{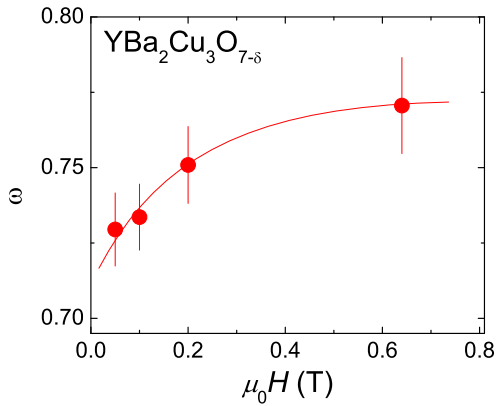}
%
\caption{(Color online) The $d-$wave contribution to the in-plane
superfluid density $\omega=\sigma^d(0)/[\sigma^s(0)+\sigma^d(0)]$
as a function of the magnetic field of
YBa$_2$Cu$_3$O$_{7-\delta}$. The line is a guide to the eye. }
 \label{fig:omega_vs_field}
\end{figure}

The $d-$wave contribution to the total superfluid density
$\omega=\sigma^d(0)/[\sigma^s(0)+\sigma^d(0)]$ increases with
increasing field (see Fig.~\ref{fig:omega_vs_field}), similar to the
field dependence observed for La$_{1.83}$Sr$_{0.17}$CuO$_4$
\cite{Khasanov07_La214}.
This dependence can be understood by the fact that superconductivity
is suppressed stronger in the $s-$wave band with increasing field
than in the $d-$wave band \cite{Khasanov07_La214}.

\begin{table}[htb]
\caption[~]{\label{Table:two-gap} Summary of the two-gap analysis
for untwined single-crystal  YBa$_2$Cu$_3$O$_{7-\delta}$ for the
magnetic field applied along the $c$~direction. The meaning of the
parameters is -- $\mu_0H$: external magnetic field, $\sigma^d(0)$
and $\sigma^s(0)$: $d-$wave and $s-$wave contribution to the
zero-temperature $\mu$SR relaxation rate $\sigma(0)$,
$\omega=\sigma^d(0)/[\sigma^s(0)+\sigma^d(0)]$: the contribution
of the large $d-$wave gap to the total in-plane superfluid
density, $\Delta^d_0$: $d-$wave gap  at $T=0$~K, $\Delta^s_0$:
$s-$wave gap at $T=0$~K. }
\begin{center}
\begin{tabular}{lcccccccc}\\ \hline
\hline
$\mu_0H$ &$\sigma^d(0)$&$\sigma^s(0)$&$\omega$&$\Delta^d_0$&$\Delta^s_0$\\
(T) &($\mu$s$^{-1}$)&($\mu$s$^{-1}$)&&(meV)&(meV)\\
\hline
0.05&1.78(2)&4.80(7)&0.729(12)&&\\
0.1 &2.01(2)&5.53(6)&0.734(11)&&\\
0.2 &1.87(2)&5.63(7)&0.751(13)&22.92(9)&0.707(11)\\
0.64&1.49(2)&5.01(7)&0.771(16)&&\\

 \hline \hline \\

\end{tabular}
   \end{center}
\end{table}

\begin{figure}[htb]
\includegraphics[width=0.9\linewidth]{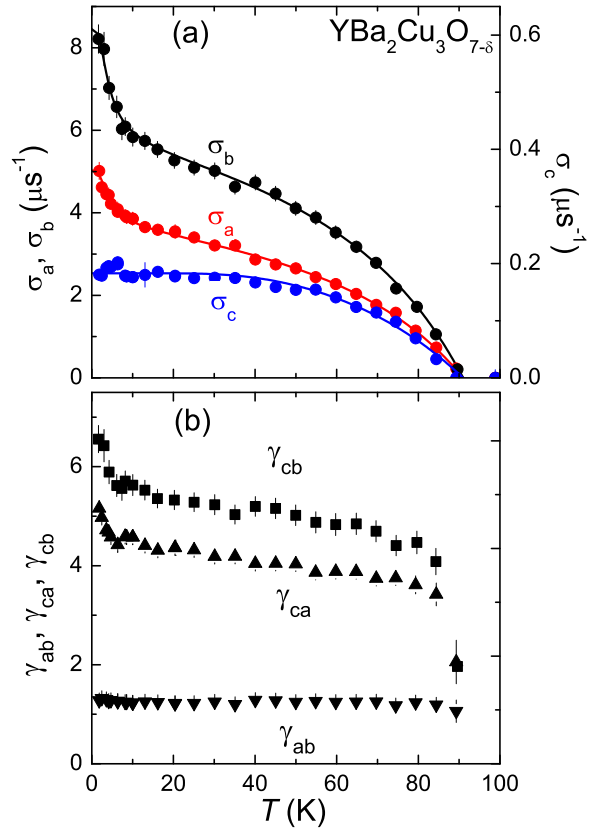}
%
\caption{(Color online) (a) Temperature dependences of
$\sigma_{a}\propto \lambda^{-2}_{a}$, $\sigma_{b}\propto
\lambda^{-2}_{b}$, and $\sigma_{c}\propto \lambda^{-2}_{c}$ of
YBa$_2$Cu$_3$O$_{7-\delta}$ obtained from $\sigma(T)$ measured
along the crystallographic $a$, $b$, and $c$~directions by using
Eq.~(\ref{eq:lambda_i}). Lines represent results of the analysis
within the two-component (black and red lines) and  one-component
(blue line) models \cite{Khasanov07_Y124}. (b) Temperature
dependences of the anisotropy parameters $\gamma_{ab}$,
$\gamma_{ca}$, and $\gamma_{cb}$ obtained as
$\gamma_{ij}=\lambda_i/\lambda_j=\sqrt{\sigma_j/\sigma_i}$ (see
text for details). }
 \label{fig:sigma_i}
\end{figure}

The individual components of the inverse squared penetration depth
($\lambda_a^{-2}$, $\lambda_b^{-2}$, and $\lambda_c^{-2}$) as
functions of temperature were obtained by applying the magnetic
field along the three crystallographic axes (0.012~T along $a$ and
$b$, and $0.1$~T along $c$). This yields the axis-related superfluid
densities according to \cite{Khasanov07_Y124}:
\begin{equation}
\sigma_i=\sigma_{ij}\cdot\sigma_{ik}/\sigma_{jk}
\propto\lambda_i^{-2}.
 \label{eq:lambda_i}
\end{equation}
The results are shown in Fig.~\ref{fig:sigma_i}~(a). Obviously
$\sigma_a$ and $\sigma_b$ have a very similar temperature
dependence, in particular, the inflection point at $T\simeq 10$~K
and the linear increase in an intermediate range of temperatures
($60$~K$\gtrsim T \gtrsim 10$~K). The temperature dependence of
$\sigma_c$ is qualitatively very different from the one of
$\sigma_a$ and $\sigma_b$. Here a saturation is observed at
$T\lesssim40$~K. $\sigma_a(T)$ and $\sigma_b(T)$ can be well
described by the two-component approach mentioned above where the
same zero-temperature gap values were used. From this analysis the
following individual contributions from the $s-$ and the $d-$wave
components along the $a$ and $b$ axis are obtained:
$\sigma_a^s(0)=1.19$~$\mu$s$^{-1}$,
$\sigma_a^d(0)=3.83$~$\mu$s$^{-1}$ and
$\sigma_b^s(0)=2.51$~$\mu$s$^{-1}$,
$\sigma_b^d(0)=5.95$~$\mu$s$^{-1}$. Since the relative contributions
of the large $d-$wave component are almost the same along $a$ and
$b$ directions, namely, $\omega_a=0.70$, $\omega_b=0.76$, it is
plausible to conclude that not the CuO chains are the cause of the
two-component behavior \cite{Atkinson95}, but that this is an
intrinsic property of cuprates. The same conclusions were reached
from different experimental techniques as, e.g., NMR
\cite{Martindale98}, Raman scattering \cite{Masui03,Friedl90}, and
ARPES \cite{Lu01}. In particular, Masui {\it et al.} \cite{Masui03}
showed that the $s+d$ symmetry of the order parameter is required in
order to describe the Raman data, even for {\it tetragonal}
Tl$_2$Ba$_2$CuO$_{6+\delta}$ HTS's.

The temperature dependence of the $c$~axis related superfluid
density $\sigma_{sc}\propto\lambda_c^{-2}$ resembles very much the
one observed for YBa$_2$Cu$_4$O$_8$ \cite{Khasanov07_Y124} and
follows closely the one expected for a single $s-$wave gap. The full
blue curve in Fig.~\ref{fig:sigma_i}~(a) corresponds to an analysis
based on an isotropic $s-$wave gap with $\Delta^s_0=17.52$~meV and
$\sigma_c(0)=0.183$~$\mu$s$^{-1}$. The present results are in good
agreement with previous findings of tunnelling experiments, where an
$s-$wave gap along the $c$~direction was reported \cite{Sun94}. The
$s-$wave component along the $c$~axis is not easily detectable by
most experimental methods because either very well oriented films
should be prepared or bulk methods have to be used. Due to the fact
that many experiments are surface sensitive only, and $ab$ oriented
samples and films are hardly available, these techniques are unable
to see the $s-$wave component along the $c$~axis. Its observation
is, however, important since the coupling of a major $d-$wave
component in the $ab$ plane to the $s-$wave component along the
$c$~axis mixes both symmetries in the planes already. In addition,
theory predicts for this scenario, that a pure $d-$wave order
parameter is never observable. On the other hand, also along the
$c$~direction an admixture of the $d-$wave order parameter has to be
present \cite{Bussmann-Holder03a}. This latter statement could
provide an explanation for the optical conductivity spectra along
the $c$~direction where strongly anisotropic gap like features have
been observed together with a finite density of states at the Fermi
energy \cite{Schutzmann94,Tajima97}.

Finally, the anisotropy along all three crystallographic directions
is addressed. This can be calculated by using
Eq.~(\ref{eq:lambda_i}) and defining the anisotropy as
$\gamma_{ij}=\lambda_i/\lambda_j=\sqrt{\sigma_j/\sigma_i}$. The
results are shown in Fig.~\ref{fig:sigma_i}~(b). While the in-plane
anisotropy ($\gamma_{ab}$) is almost constant for all temperatures,
both out-of-plane components ($\gamma_{ca}$ and $\gamma_{cb}$)
exhibit a sharp increase,
as expected from Fig.~\ref{fig:sigma_i}~(a). In addition, it is seen
that while $\gamma_{ab}$ is almost always close to 1.2,
$\gamma_{ca}$, $\gamma_{cb}$ are substantially larger and  reach
values between 5 and 6.5 in the low-$T$ limit. This high anisotropy
is in very good agreement with previously reported values
\cite{Ishida97,Ager00}. It reflects the fact that the CuO$_2$ planes
have nearly Fermi liquid-like metallic properties whereas along the
$c$~direction mostly insulating behavior is observed.

Our conclusions from the above presented data are manifold. Since
$s+d-$wave symmetries of the superconducting order parameter were
observed previously in various cuprate families by various different
techniques \cite{Martindale98,Masui03,Friedl90,Smilde05,
Deutscher05,Lu01,Khasanov07_La214,Khasanov07_Y124,
Khasano07_La214_book,Furrer07}, the new $\mu$SR data together with
earlier results on structurally different compounds
\cite{Khasanov07_La214,Khasanov07_Y124, Khasano07_La214_book}
support the idea that this behavior is {\it intrinsic} and {\it
universal}. Similarly, the observation of an $s-$wave order
parameter along the crystallographic $c$~axis is proposed to be
intrinsic as well. Specifically, this latter point emphasizes the
importance of the {\it third} dimension for HTS's which was
neglected in most of the theoretical models. In this context it is
worth mentioning that the importance of the $c$~axis has already
been emphasized from ab-initio band structure calculations, where
trends in $T_c$ were correlated with CuO$_2$ apical oxygen distances
\cite{Pavarini01}. Also, from first principles doping dependent
computations of ARPES intensities, it was concluded that
contributions from the $c$~axis are of crucial importance in
understanding the physics of HTS's \cite{Sahrahorpi05}. On the other
hand, the observation of mixed order parameters and more
specifically, the additional $s-$wave component, require that the
lattice must be considered in the physics of HTS's.

It is a pleasure to acknowledge many stimulating and supporting
discussions with K. A. M\"uller. This work was partly performed at
the Swiss Muon Source (S$\mu$S), Paul Scherrer Institute (PSI,
Switzerland). The authors are grateful to A.~Amato and D.~Herlach
for assistance during the $\mu$SR measurements. This work was
supported by the Swiss National Science Foundation, by the
K.~Alex~M\"uller Foundation, and the EU Project CoMePhS.

\end{document}